\documentclass[aps,showpacs,pre,superscriptaddress,preprint]{revtex4}%
\usepackage{amsfonts}
\usepackage{amsmath}
\usepackage{amssymb}
\usepackage{wasysym}
\usepackage{graphicx}
\usepackage{color}%
\newcommand{\placefigure}[4]
{
\begin{figure}[t]
\includegraphics[width=#2]{#1}
\caption{#3} \label{#4}
\end{figure}
}

\begin{document}

\title{QUANTUM ALGEBRA IN THE MIXED LIGHT PSEUDOSCALAR MESON STATES}

\author{Li-Jun Tian}
\email{tianlijun@shu.edu.cn}
\affiliation{Department of Physics,
Shanghai University, Shanghai 200444, P.R. China}
\author{Yan-Ling Jin}
\email{jinyanling@shu.edu.cn}
\affiliation{Department of Physics,
Shanghai University, Shanghai 200444, P.R. China}
\author{Ying Jiang}
\email{yjiang@shu.edu.cn}
\affiliation{Department of Physics,
Shanghai University, Shanghai 200444, P.R. China}

\begin{abstract}
In this paper, we investigate the entanglement degrees of
pseudoscalar meson states via quantum algebra $Y(su(3))$. By
making use of transition effect of generators $J$ of $Y(su(3))$,
we construct various transition operators in terms of $J$ of, and
act them on $\eta$-$\pi^0$-$\eta'$ mixing meson state. The
entanglement degrees of both the initial state and final state are
calculated with the help of entropy theory. The diagrams of
entanglement degrees are presented. Our result shows that a state
with desired entanglement degree can be achieved by acting proper
chosen transition operator on an initial state. This sheds new
light on the connect among quantum information, particle physics
and Yangian algebra.
\end{abstract}


\maketitle

\section{Introduction}

It has been realized that quantum entanglement is a key ingredient
in quantum computation, quantum communication, and quantum
cryptography\cite{Galindo}. Many two-level quantum systems, or
qubits, have been widely used in quantum
information\cite{Deng,Lu}. Recently, the bipartite qutrit systems
have drawn people's attention and exhibited varieties of
advantages. It enables powerful computation\cite{Andrew},
establishes secure communication\cite{Langford} and
cryptography\cite{Pasquinucci}, and reduces the communication
complexity\cite{Brukner}. Additionally, the high energy quantum
teleportation using neutral kaons has been investigated\cite{Shi}.
These hint the possibility of connecting quantum information and
particle physics, this connection may reveal novel and interesting
feature. Considering these two aspects, we take the mixed light
pseudoscalar meson states, a type of bipartite qutrit system, into
account.

In particle physics, the study on mixed meson
states\cite{gellmann-mixed} plays an important role and many works
are devoted to this field\cite{feldmann-kroll,magiera,Kroll},
especially the study on $\eta-\eta^{'}$ mixed state, since it
provides a unique opportunity for testing QCD \cite{Dmitrasinovic}
which is widely used in describing strong interaction. In the last
decades, people have studied the mixing angle\cite{Cao}, the
hadronic $\eta$-$\eta'$ decay\cite{Borasoy}, $\eta$-$\eta'$ mixing
in radiative $\phi$-meson decay\cite{Fazio}, and so on. However,
much work of particle physics has been done for testing the theory
and experiment while hardly concerned with quantum information and
the related algebra configuration, such as Yangians.  As is known,
the light pseudoscalar meson states in quark-flavor basis have
$su(3)$ symmetry\cite{Han}. It is important to expand the research
on the symmetry of pseudoscalar meson states to Yangian
$Y(su(3))$.

Yangian, as an algebra beyond the Lie algebra, is a powerful
mathematical method for investigating the new symmetry of quantum
systems which are nonlinear and integrable. Demonstrating and
investigating whether simple physical systems possess Yangian is
important and helpful for exploring physical systems via quantum
algebra. People have found the Yangian symmetry in many physical
models, such as Calogero-Sutherland model\cite{Uglov}, the Hubbard
model\cite{Kundu} and the Heisenberg model\cite{bernard2}, etc.
The realizations of $Y(sl(2))$\cite{Haldane,Shastry} as the
simplest one in Yangian algebra have been gained much attention.
However, the Yangian related to the Lie algebra $su(3)$,
$Y(su(3))$, which is closed to the light pseudoscalar meson states
in particle physics, need to be investigated in more detail. Thus,
much attention is going to be payed in this paper to the
application of $Y(su(3))$ algebra in the meson systems here.

\section{\label{sec1}$Y(su(3))$ Algebra In $\eta$-$\pi^0$-$\eta'$ Mixing System}

In the low mass hadron region, the violations of isospin symmetry
for pseudoscalar meson states within QCD are generated by the
admixtures of $\eta$-$\pi^0$-$\eta'$. $\eta$ and $\eta'$ are
linear combination of $su(3)$ singlet $\eta^{0'}$ and octet
$\eta^0$, $\pi^0$ is another $su(3)$ octet. Because of the
important application and significance of $\eta$-$\pi^0$-$\eta'$
mixing, we choose their superposition states as the initial state
\begin{eqnarray}
|\phi\rangle=\alpha_1|\eta^{0'}\rangle+\alpha_2|\pi^0\rangle+\alpha_3|\eta^0\rangle,
\label{eq1}
\end{eqnarray}
where $\alpha_1$, $\alpha_2$ and $\alpha_3$ are the normalized real
amplitudes and they satisfy $\alpha_1^2+\alpha_2^2+\alpha_3^2=1$.
$|\eta^{0'}\rangle=\frac{\sqrt3}{3}(|u\bar{u}\rangle+|d\bar{d}\rangle+|s\bar{s}\rangle)$,
$|\eta^0\rangle=\frac{\sqrt6}{6}(-|u\bar{u}\rangle-|d\bar{d}\rangle+2|s\bar{s}\rangle)$,
$|\pi^0\rangle=\frac{\sqrt2}{2}(|u\bar{u}\rangle-|d\bar{d}\rangle)$.

The entanglement degree of the genuine N-particle qutrit pure
state\cite{Pan} is measured by its mean entropy
\begin{eqnarray}
\label{2} C^{(N)}_\Phi=\left\{
\begin{array}{l}
\frac1N\sum_{i=1}^nS_{(i)}\;\;\;\;\;$if$\;S_i\neq0\;\forall\;i\\
0\;\;\;\;\;\;\;\;\;\;\;\;\;\;\;\;\;\;\;\;\;\;$otherwise$
\end{array} \right.,
\end{eqnarray}
where $S_i=-Tr((\rho_\Phi)_iLog_3(\rho_\Phi)_i)$ is the reduced
partial Von Neumann entropy for the $i$th particle only, with the
other $N-1$ particles traced out, and $(\rho_\Phi)_i$ is the
corresponding reduced density matrix. Since the system in Eq.
(\ref{eq1}) is bipartite qutrit, i.e. $N=2$, the corresponding
entanglement degree of this initial state can be calculated via
Eq.(\ref{2}), thus we have
\begin{eqnarray}
 C_\phi =-(\frac{\sqrt3}{3}\alpha_1+\frac{\sqrt2}{2}\alpha_2-\frac{\sqrt6}{6}\alpha_3)^2Log_3
 (\frac{\sqrt3}{3}\alpha_1+\frac{\sqrt2}{2}\alpha_2-\frac{\sqrt6}{6}\alpha_3)^2\nonumber\\
 -(\frac{\sqrt3}
 {3}\alpha_1-\frac{\sqrt2}{2}\alpha_2-\frac{\sqrt6}{6}\alpha_3)^2Log_3(\frac{\sqrt3}{3}\alpha_1
 -\frac{\sqrt2}{2}\alpha_2-\frac{\sqrt6}{6}\alpha_3)^2\nonumber\\
 -(\frac{\sqrt3}{3}\alpha_1+\frac{\sqrt6}
 {3}\alpha_3)^2Log_3(\frac{\sqrt3}{3}\alpha_1+\frac{\sqrt6}{3}\alpha_3)^2.
  \end{eqnarray}
The behavior of $C_\phi$ depending on $\alpha_1$ and $\alpha_2$ (
due to the normalization condition, $\alpha_3$ is not an
independent parameter.) is given in Fig.~\ref{fig1}.

\placefigure{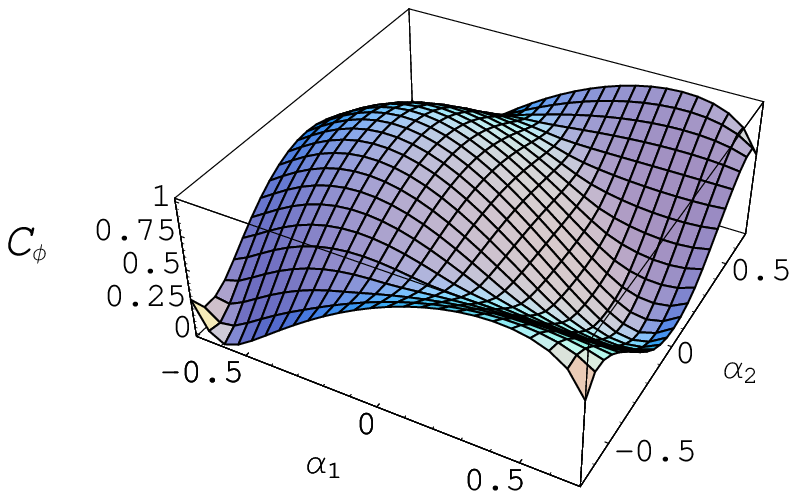}{0.4\linewidth}{(color online). The entanglement
degree of the initial state $|\phi\rangle$ for different
$\alpha_1$ and $\alpha_2$. }{fig1}

Before constructing transition operators, let us first give a
brief introduction of Yangian $Y(su(3))$. $Y(su(3))$
algebra\cite{Ge} is generated by the generators $\{{I}^a,{J}^a\}$
which are usually defined as follows
\begin{eqnarray}
&&I^a=\sum_iF_i^a,\nonumber\\
  &&J^a=\mu{I_1^a}+\nu{I_2^a}+\frac{i}{2}\lambda{f_{abc}\sum_{i{\neq}j}{\omega}_{ij}I_i^bI_j^c}\;\;(i,j=1,2).\;\;\;
 \end{eqnarray}
Here $I^a$ form a $su(3)$ algebra characterized by $f_{abc}$,
$\{F_{i}^{a}, a=1,\cdots,8\}$ form a local $su(3)$ on the $i$
site, and are equal to half of the corresponding Gell-Mann
matrices, $\mu$, $\nu$, $\lambda$ are parameters or Casimir
operators and $\omega_{ij}=-\omega_{ji}$ which satisfies
\begin{equation}
{\omega}_{ij}=\left\{
\begin{array}{l}
1\;\;\;\;\;\;\;\;i{>}j\\
-1\;\;\;\;\;i{<}j\\
0\;\;\;\;\;\;\;\;i{=}j
\end{array} \right. .
\end{equation}

A more practical expression is expressed as
\begin{eqnarray}
{\bar{{I}}^{\pm}}=J^{1}{\pm}iJ^{2},\;\;\;\;{\bar{U}^{\pm}}=J^{6}{\pm}iJ^{7},\;\;\;\;
{\bar{V}^{\pm}}=J^{4}{\pm}iJ^{5},\;\;\;\;
{\bar{I}^{3}}=J^{3},\;\;\;\; {\bar{I}^{8}}=\frac{2}{\sqrt{3}}J^{8}.
\label{a}
\end{eqnarray}
Here $J^a\;(a =1,\cdots,8)$ are the generators of $Y(su(3))$.

Due to the transition effect of Yangian generators, transition
operators $P$ can be constructed as compositions of the generators
in Eq.~(\ref{a}). $P$ can be looked upon as a function of
${\bar{{I}}^{\pm}}, {\bar{U}^{\pm}}, {\bar{V}^{\pm}},
{\bar{I}^{3}}$ and ${\bar{I}^{8}}$, namely,
$P=F[{\bar{{I}}^{\pm}}, {\bar{U}^{\pm}}, {\bar{V}^{\pm}},
{\bar{I}^{3}}, {\bar{I}^{8}}]$. When acting the transition
operator on $|\phi\rangle$ in Eq.~(\ref{eq1}), a final state
$|\phi^{'}\rangle=P|\phi\rangle$ can be gotten and its
entanglement degree $C_{\phi^{'}}$ can also be calculated out via
Eq.(\ref{2}). Different final states with desired entanglement
degrees thus can be gotten by acting corresponding transition
operators on initial states.

In order to illustrate this issue clearly, several simple examples
are going to be discussed in more detail.

\section{\label{sec5}Obtaining different entanglement degrees by
tuning transition operators}

As we discussed before, the entanglement degree of final state can
be tuned by changing the parameters of the transition operator. As
an example, let us act the transition operator of
$P=\bar{V}^++\bar{V}^-$ on the initial state in Eq.(\ref{eq1}),
the corresponding final state is
\begin{eqnarray}
|\phi^{'}\rangle=P|\phi\rangle=(\nu+\frac{\lambda}{2})[(\frac{\sqrt3}{3}\alpha_1+\frac{\sqrt2}{2}\alpha_2-
\frac{\sqrt6}{6}\alpha_3)|K^+\rangle+(\frac{\sqrt3}{3}\alpha_1+
\frac{\sqrt6}{3}\alpha_3)|K^-\rangle].
\end{eqnarray}
The normalizing condition leads to $\mu=\frac{\lambda}{2}$ and $
(\nu+\frac{\lambda}{2})^2[1-(\frac{\sqrt3}{3}\alpha_1-\frac{\sqrt2}{2}\alpha_2-
\frac{\sqrt6}{6}\alpha_3)^2]=1$. As discussed in the preceding
section, the entanglement degree of the final state
$|\phi^{'}\rangle$ can be calculated by the use of Eq.(\ref{2}) as
\begin{eqnarray}
C_{\phi^{'}}=-(\nu+\frac{\lambda}{2})^2(\frac{\sqrt3}{3}\alpha_1+\frac{\sqrt2}{2}\alpha_2-
\frac{\sqrt6}{6}\alpha_3)^2Log_3(\nu+\frac{\lambda}{2})^2(\frac{\sqrt3}{3}\alpha_1+\frac{\sqrt2}{2}\alpha_2-
\nonumber\\
\frac{\sqrt6}{6}\alpha_3)^2-(\nu+\frac{\lambda}{2})^2(\frac{\sqrt3}{3}\alpha_1+
\frac{\sqrt6}{3}\alpha_3)^2Log_3(\nu+\frac{\lambda}{2})^2(\frac{\sqrt3}{3}\alpha_1+
\frac{\sqrt6}{3}\alpha_3)^2.
\end{eqnarray}
As we can see, by changing the value of $\nu$ and $\lambda$, the
entanglement degree $C_{\phi'}$ can be tuned. This behavior is
shown in Fig.~\ref{fig3} for the case of
$\alpha_1=\alpha_2=\frac12$.

\placefigure{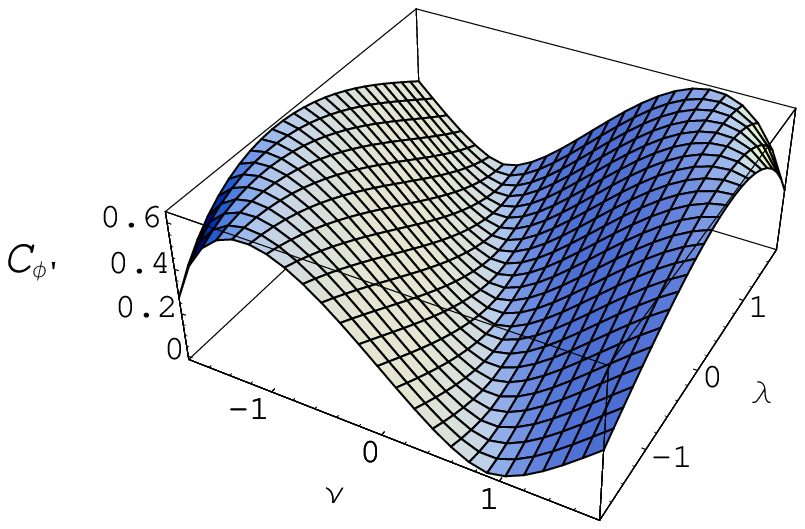}{0.4\linewidth}{(color online). The entanglement
degree of the final state $|\phi^{'}\rangle$ varies with $\nu$ and
$\lambda$ when $\alpha_1=\alpha_2=\frac12$.}{fig3}

Another more simpler case is $P=\bar{I}^8$ for which the final
state reads
\begin{eqnarray}
|\phi^{'}\rangle=P|\phi\rangle=\frac13(\nu+\frac{\lambda}{2})[-\frac{\sqrt2+2\sqrt6}{3}\alpha_3|\eta^{0'}\rangle+\alpha_2|\pi^0\rangle-(\sqrt2\alpha_1+\alpha_3)|\eta^0\rangle]
\end{eqnarray}
with normalization condition $\mu=\frac{\lambda}{2}$ and $
(\nu+\frac{\lambda}{2})^2[1+3(\frac{\sqrt3}{3}\alpha_1+
\frac{\sqrt6}{3}\alpha_3)^2]=3$. Similarly,
\begin{eqnarray}
C_{\phi^{'}}=-\frac19(\nu+\frac{\lambda}{2})^2(\frac{\sqrt3}{3}\alpha_1+\frac{\sqrt2}{2}\alpha_2-
\frac{\sqrt6}{6}\alpha_3)^2Log_3\frac19(\nu+\frac{\lambda}{2})^2(\frac{\sqrt3}{3}\alpha_1+\frac{\sqrt2}{2}\alpha_2-\nonumber\\
\frac{\sqrt6}{6}\alpha_3)^2-\frac19(\nu+\frac{\lambda}{2})^2(\frac{\sqrt3}{3}\alpha_1-\frac{\sqrt2}{2}\alpha_2-
\frac{\sqrt6}{6}\alpha_3)^2Log_3\frac19(\nu+\frac{\lambda}{2})^2(\frac{\sqrt3}{3}\alpha_1-\frac{\sqrt2}{2}\alpha_2\nonumber\\
-\frac{\sqrt6}{6}\alpha_3)^2-\frac49(\nu+\frac{\lambda}{2})^2(\frac{\sqrt3}{3}\alpha_1+
\frac{\sqrt6}{3}\alpha_3)^2Log_3\frac49(\nu+\frac{\lambda}{2})^2(\frac{\sqrt3}{3}\alpha_1+
\frac{\sqrt6}{3}\alpha_3)^2,
\end{eqnarray}
which also depends on $\nu$ and $\lambda$, as shown in
Fig.~\ref{fig5}.

\placefigure{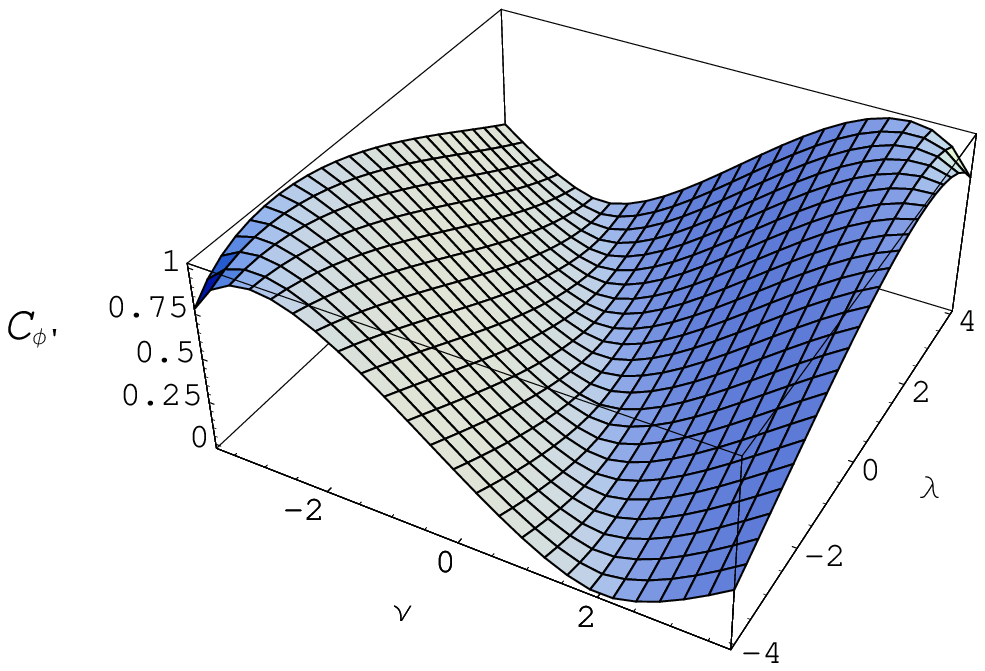}{0.4\linewidth}{(color online). The entanglement
degree of the final state $|\phi^{'}\rangle$ varies with $\nu$ and
$\lambda$ when $\alpha_1=\alpha_2=\frac12$.}{fig5}

These two examples clearly shows that by tuning the parameters
$\nu$ and $\lambda$ of transition operators, the entanglement
degree of final state varies between 0 and 1.

However, there are exceptional cases in which the entanglement
degree of final state is found to be the same as the initial state
or be zero, independent with $\nu$ and $\lambda$, as is shown in
the next section.

\section{\label{sec3} Two exceptional cases }

Although generally, transition operators with different $\nu$ and
$\lambda$ provide final states with different entanglement
degrees, there do exist special cases in which the final state
possesses the same entanglement degree with the initial state or
the final state are totally disentangled. Both cases are very
important in quantum information and quantum computation.

For the transition operator $P=\bar{I}^-+\bar{U}^-+\bar{V}^-$, the
final state is
\begin{eqnarray}
|\phi^{'}\rangle=P|\phi\rangle=-(\frac{\sqrt3}{3}\alpha_1+\frac{\sqrt2}{2}\alpha_2-
\frac{\sqrt6}{6}\alpha_3)|\pi^+\rangle+(\frac{\sqrt3}{3}\alpha_1-
\frac{\sqrt2}{2}\alpha_2-\frac{\sqrt6}{6}\alpha_3)|K^0\rangle+\nonumber\\
(\frac{\sqrt3}{3}\alpha_1+\frac{\sqrt6}{3}\alpha_3)|K^-\rangle
\end{eqnarray}
with normalizing condition $\mu+\nu=1$ and
$\mu=\frac{\lambda}{2}$. It is very easy to verify that the
entanglement degree of the final state $|\phi^{'}\rangle$ is equal
to the one of the initial state, namely, $C_{\phi^{'}}=C_\phi$.

Same thing happens on transition operator
$P=\bar{I}^++\bar{U}^++\bar{V}^+$, corresponding to whom the final
state reads
\begin{eqnarray}
|\phi^{'}\rangle=P|\phi\rangle=-(\frac{\sqrt3}{3}\alpha_1+\frac{\sqrt2}{2}\alpha_2-
\frac{\sqrt6}{6}\alpha_3)|K^+\rangle+(\frac{\sqrt3}{3}\alpha_1-
\frac{\sqrt2}{2}\alpha_2-\frac{\sqrt6}{6}\alpha_3)|\pi^-\rangle\nonumber\\
+(\frac{\sqrt3}{3}\alpha_1+\frac{\sqrt6}{3}\alpha_3)|\bar{K}^0\rangle
\end{eqnarray}
with normalizing condition being $\mu+\nu=1$ and
$\mu=\frac{\lambda}{2}$. Again, calculation shows that
$C_{\phi^{'}}=C_\phi$.

Without any difficulty, it can be verified that transition
operators $P=\bar{I}^-+\bar{V}^+$, $P=\bar{I}^++\bar{U}^-$, and
$P=\bar{U}^++\bar{V}^-$ lead to total disentangled final states,
independent of the choice of $\alpha_1$ and $\alpha_2$ in the
initial state.

\section {\label{sec2}Conclusions}

In conclusion, we have investigated the entanglement degrees of
pseudoscalar meson states via quantum algebra $Y(su(3))$. By
making use of transition effect of generators $J$ of $Y(su(3))$,
we have constructed various transition operators in terms of $J$,
and have acted them on $\eta$-$\pi^0$-$\eta'$ mixing meson state.
The entanglement degrees of both the initial state and final state
have been calculated with the help of entropy theory. Our result
shows that a state with desired entanglement degree can be
achieved by acting proper chosen transition operator on an initial
state. This is very helpful to control the degree of entanglement
in quantum communication. Entanglement has been considered as an
essential resource in most applications of quantum
information\cite{Galindo}. Although the quantum channels used in
quantum teleportation are usually represented by a maximally
entangled pair\cite{Bennett}, partially entangled quantum channel
becomes a hot topic\cite{Fang} nowadays due to the noise factors
in the realistic world. Moreover, in quantum computing, a new
important development is the probabilistic implementation of a
nonlocal gate by using a single non-maximally entangled
state\cite{Chen}. The success fidelity maintains a high level
though the state shared is partially entangled.

We believe that our work provide a connection among particle
physics, quantum information and quantum algebra, and the result
may shed new light on entanglement controlling in quantum
computing.


\begin{thebibliography}{99}
\bibitem{Galindo} A. Galindo and M. A. M. Delgado, {\it Rev. Mod.
Phys.} {\bf{74}} (2002) 347.
\bibitem{Deng}F. G. Deng, G. L. Long and X. S. Liu, {\it Phys. Rev. A}
{\bf{68}} (2003) 042317.
\bibitem{Lu}H. Lu and G. C. Guo, {\it Phys. Lett. A} {\bf{276}}
(2000) 209.
\bibitem{Andrew}A. M. Childs and I. L. Chuang, {\it Phys. Rev. A}
{\bf{63}} (2000) 012306.
\bibitem{Langford} N. K. Langford, R. B. Dalton, M. D. Harvey, J. L. \'{O}Brien, G. J. Pryde, A. Gilchrist, S. D. Bartlett and A. G. White, {\it Phys.
Rev. Lett.} {\bf{93}} (2004) 053601.
\bibitem{Pasquinucci}H. B. Pasquinucci and A. Peres, {\it Phys. Rev.
Lett.} {\bf{85}} (2000) 3313.
\bibitem{Brukner}\v{C}. Brukner, M. \.{Z}ukowski and A. Zeilinger, {\it Phys.
Rev. Lett.} {\bf{89}} (2002) 197901.
\bibitem{Shi}Y. Shi, {\it Phys. Lett. B} {\bf{641}} (2006) 75.
\bibitem{gellmann-mixed}M.Gell-Mann and A. Pais, Phys. Rev. 97 (1955) 1387.
\bibitem{feldmann-kroll}T. Feldmann and P. Kroll, Phys. Rev. D 58 (1998) 114006
\bibitem{magiera}A. Magiera and H. Machner, Nucl. Phys. A 674 (2000) 515
\bibitem{Kroll}P. Kroll, {\it Modern Phys. Lett. A} {\bf{20}} (2005) 2667.
\bibitem{Dmitrasinovic}V. Dmitrasinovic, {\it Phys. Rev.D} {\bf{56}}
(1997) 247.
\bibitem{Cao}J. Cao, F. G. Cao, T. Huang and B. Q. Ma, {\it Phys. Rev. D} {\bf{58}} (1998) 113006.
\bibitem{Borasoy} B. Borasoy and R. Nilbler, {\it Phys. J. A}
{\bf{26}} (2005) 383.
\bibitem{Fazio} F. D. Fazio and M. R. Pennington, {\it J. High Energy Phys.} {\bf{7}} (2000) 51.
\bibitem{Han} M. Y. Han and Y. Nambu, {\it Phys. Rev.} {\bf{139}} (1965) B1006.
\bibitem{Uglov} D. Uglov, {\it Commun. Math. Phys.} {\bf{191}} (1998) 663.
\bibitem{Kundu} A. Kundu, {\it Phys. Lett. A} {\bf{249}} (1998) 126.
\bibitem{bernard2} D. Bernard, {\it Inter. J. Modern Phys. B} {\bf{7}} (1993) 3517.
\bibitem{Haldane} F. D. M. Haldane, {\it Phys. Rev. Lett.}
{\bf{60}} (1988) 635.
\bibitem{Shastry} B. S. Shastry, {\it Phys. Rev. Lett.} {\bf{60}} (1988) 639.
\bibitem{Pan} F. Pan, D. Liu, G. Y. Lu and J. P. Draayer, {\it Phys. Lett. A} {\bf{336}} (2005) 384.
\bibitem{Ge} M. L. Ge. and K. Xue, {\it Yang$-$Baxter Equation} (Shanghai: Shanghai Scientific and Technical
Publishers) (1999) p511.
\bibitem{Bennett} C. H. Bennett, G. Brassard, C. Cr\'{e}peau, R. Jozsa, A. Peres and W. K. Wootters, {\it Phys. Rev.
Lett.} {\bf{70}} (1993) 1895.
\bibitem{Fang} J. X. Fang, Y. S. Lin, S. Q. Zhu and X. F. Chen, {\it Phys. Rev. A}
{\bf{67}} (2004) 014305.
\bibitem{Chen} L. Chen and Y. X. Chen, {\it Phys. Rev. A} {\bf{71}} (2005) 054302.

\end{thebibliography}
\end{document}